\documentclass[%
 reprint,
 amsmath,amssymb,
 aps,
prl,
]{revtex4-2}
\usepackage{xcolor}
\usepackage{graphicx}
\usepackage{dcolumn}
\usepackage{bm}

\newcommand{\apjl}{{Astrophys. J. Lett.}}
\newcommand{\araa}{{Annual Review of Astronomy and Astrophysics}}
\newcommand{\apss}{{Astrophysics and Space Science}}
\newcommand{\mnras}{{Monthly Notices of the Royal Astronomical Society}}
\newcommand{\pasp}{{Publications of the Astronomical Society of the Pacific}}
\begin{document}

\preprint{APS/123-QED}

\title{The anisotropy and magnetic field structure of neutron stars through gravitational wave}

\author{Zhao-Wei Du}
\author{Xi-Long Fan} 
 \email{xilong.fan@whu.edu.cn}
\affiliation{School of Physics Science and Technology, Wuhan University, No.299 Bayi Road, Wuhan, Hubei, China}

\date{\today}

\begin{abstract}
We investigate how gravitational wave (GW) observations can probe the internal physics of neutron stars by extending the Tolman–Oppenheimer–Volkoff framework to include pressure anisotropy and internal magnetic fields. Two representative magnetic field configurations, radial orientation dominated (RO) and transverse orientation dominated (TO), are implemented with strength and decay prescriptions. We found that both anisotropy and magnetic fields increase the maximum supported mass and modify the tidal deformability $\Lambda$, thereby imprinting measurable signatures on GW signals. For the equal mass binary ($1.2M_\odot$–$1.2M_\odot$), anisotropy neutron star with RO magnetic field yield more compact stars and a larger shift in $\Lambda$, allowing discrimination at signal-to-noise ratios (SNRs) as low as $\sim18$ using the O4 power spectra density. TO fields produce weaker effects and require substantially higher SNRs for detection. In conclusion, we conclude that gravitational waves are capable of probing the internal structure of neutron stars.
\end{abstract}

\maketitle


\section{INTRODUCTION}
Neutron stars are among the most popular compact objects in the Universe. Their surface magnetic fields can reach $\sim10^{15}~{\rm G}$, while the interior may sustain fields as large as $\sim10^{18}~{\rm G}$; central densities exceed $\sim10^{15}~{\rm g~cm^{-3}}$ \cite{1998A&A...335..969Y, 2014PhRvC..89d5805M, 2015RPPh...78k6901T}. These extremes conditions make neutron stars powerful engines for both electromagnetic (EM) and GW emission, expecially in binary neutron-star (BNS) mergers \cite{2017PhRvL.119p1101A, 2017ApJ...848L..14G, 2017Natur.551...67P, 2018NatCo...9..447Z, 2025ApJ...985...42D}, turning them into cosmic laboratories for physics beyond terrestrial reach. However, the interior physics of neutron stars, particularly the magnetic field structure, strength of anisotropy and the equation of state, is not yet tightly constrained. Uncertainty persists because dense matter microphysics is hard to compute beyond nuclear saturation, and observations suffer model/systematic degeneracies. Precisely determining the nature would pin down neutron star radii, maximum masses, and tidal deformabilities. In turn, it would yield more accurate estimates from BNS mergers and other neutron star phenomena, substantially advancing constraints on dense-matter microphysics and potential new physics. 


Since terrestrial experiments cannot access such extreme densities, the study of neutron star physics is constrained predominantly by EM and GW observations. A particularly effective GW probe of the equation of state (EoS) is the mass–tidal deformability relation $M_{\rm NS}$–$\Lambda$, extracted from the tidal phase corrections in binary neutron-star inspirals \cite{2008PhRvD..77b1502F, 2009PhRvD..79l4032R, 2018PhRvL.121i1102D}. The tidal deformability is the star’s linear, quadrupolar response to an external tidal field \cite{2008ApJ...677.1216H}, reflecting the nature of the EoS and neutron star structure. GW measurements currently yield some of the tightest model-independent constraints. The landmark event GW170817 yields $\Lambda_{1.4}\leq800$ ($\leq1400$ under a high-spin prior) for a neutron star with $1.4~M_\odot$ \cite{2017PhRvL.119p1101A}, subsequent analyses using refined waveform models and equation-of-state constraints have tightened this bound to $\Lambda_{1.4} \approx 190_{-120}^{+390}$ (approximating an upper limit of $\sim 580$) \cite{2018PhRvL.121p1101A,2019PhRvX...9a1001A}.

Anisotropy and magnetic fields may play important roles in neutron stars. Most analyses simplely assume that the neutron star is pressure isotropy. However, many studies argued that an anisotropic star may be more realistic \cite{1972ARA&A..10..427R, 1974ApJ...188..657B, 1980PhRvD..22..807L, 1982PhRvD..26.1262B, 1993Ap&SS.201..191B, 1998NuPhB.531..478C}. Bowers \& Liang \cite{1974ApJ...188..657B} proposed that pressure anisotropy can have non-negligible effects on neutron-star properties, including changes to the mass–radius relation for a fixed EoS. Many papers have argued that strong magnetic fields can influence neutron stars, for example, by stiffening or softening the EoS and introducing extra pressure arising from Lorentz force \cite{2000ApJ...537..351B, 2001ApJ...554..322C, 2010PhRvC..82b5804R, 2014PhRvC..89a5805C, 2015ChPhC..39a5101H, 2020EPJA...56....2K}. Both the orientation and the magnitude of magnetic field have a significant effect on neutron star \cite{2014PhRvD..90f3013C}. The combination of anisotropy and magnetic field cast a veil of mystery over neutron star. Several studies aim to infer the internal distributions of anisotropic pressure and magnetic fields in neutron stars \cite{2021ApJ...922..149D, 2024PhRvD.109b3027Z}. In particular, the magnetic field structure inside crust may be linked to electromagnetic activity, such as fast radio bursts (FRBs) and short X-ray bursts \cite{2019MNRAS.488.5887S, 2019ApJ...879....4W, 2021ApJ...919...89Y, 2023MNRAS.526.2795T}. Attempts to constrain the magnetic field within the neutron star crust have primarily relied on electromagnetic observables, most notably quasi-periodic oscillations (QPOs) in soft gamma-ray repeaters (SGRs) and thermal evolution modeling \cite{2005ApJ...628L..53I, 2012MNRAS.421.2054G, 2013PhRvL.111u1102G, 2018MNRAS.476.4199G, 2008ApJ...673L.167A, 2013MNRAS.434.2362P}. However, as GW detector sensitivity continues to improve, GWs offer a complementary, independent avenue for constraining the internal magnetic field distribution in neutron stars. A systematic investigation of the capacity of GW detectors to distinguish these effects is therefore indispensable, notably for current instruments.

In this paper, we explore how GWs can be used to probe the internal pressure anisotropy and magnetic field structure of neutron stars. In Section 2, we summarize the TOV equations modified to include magnetic fields and anisotropic pressure, review the calculation of tidal deformability in the presence of anisotropy, and derive updated $M_{\rm NS}-\Lambda$ and $M_{\rm NS}-R_{\rm NS}$ relations over a range of anisotropy and magnetic-field parameters. In Section 3, we quantify the SNR required to distinguish between the standard TOV solutions and the modified TOV models throughout the parameter space. In Section 4, we summarize our main results and discuss their implications for identifying the effects of internal magnetic fields and pressure anisotropy in gravitational-wave observations. The EoS adopted in this work is SLy4 \cite{1998NuPhA.635..231C, 2009NuPhA.818...36D, 2015PhRvC..92e5803G}, obtained from the CompOSE database (https://compose.obspm.fr
) \cite{2015PPN....46..633T, 2017RvMP...89a5007O, 2022EPJA...58..221C}. Throughout the paper, we use geometrized units with $G=c=1$, while numerical values are quoted in cgs units when appropriate.

\section{MODIFIED TOV AND TIDAL DEFORMABILITY}
\subsection{Modified TOV Equation}
The metric of anisotropical neutron star is assume to be static and spherically symmetric, hence we set
\begin{equation}
    \mathrm{d} s^{2}=e^{\nu(r)} \mathrm{d} t^{2}-e^{\lambda(r)} \mathrm{d} r^{2}-r^{2}\left(\mathrm{d} \theta^{2}+\sin ^{2} \theta \mathrm{d} \phi^{2}\right),\label{metric}
\end{equation}
where $\nu(r)$ and $\lambda(r)$ are functions of the radial coordinate $r$.

To derive the TOV equations modified by magnetic fields and pressure anisotropy, we need to construct the total energy-momentum tensor \cite{2021ApJ...922..149D}
\begin{equation}
    T^{\mu\nu}  = T_m^{\mu\nu} + T_f^{\mu\nu}\label{total em tensor}
\end{equation}
with
\begin{equation}
\begin{aligned}
    T_m^{\mu\nu} = (&\rho + p_t) u^\mu u^\nu - p_t g^{\mu\nu} + (p_r - p_t) v^\mu v^\nu, \\
    T_f^{\mu\nu} &= \frac{B^2}{4\pi} \left( u^\mu u^\nu - \frac{1}{2} g^{\mu\nu} \right) - \frac{B^\mu B^\nu}{4\pi},
    \end{aligned}
\end{equation}
where $T_m^{\mu\nu}$ and $T_f^{\mu\nu}$ denote energy-momentum tensors of matter and magnetic field, respectively. $\rho$, $p_r$, and $p_t$ denote the energy density, the radial pressure, and the transverse pressure, respectively, with $p_t$ defined orthogonal to the radial direction. $B^2=-B^\mu B_\mu$ represents the squared magnitude of the magnetic field. In a static, spherically symmetric spacetime we take
$$
u^\mu =\delta^\mu{}_0 e^{-\nu(r)/2},\qquad
v^\mu = \delta^\mu{}_1e^{-\lambda(r)/2},
$$
so that $u^\mu$ is orthogonal to $v^\mu$, i.e. $u_\mu v^\mu = 0$. With the $(+,-,-,-)$ signature they satisfy $u_\mu u^\mu=-v_\mu v^\mu=1$. We negelect the term of magnetization in $T_m^{\mu\nu}$ due to they are at least an order of magnitude smaller than magnetic pressure and has no appreciable effect on magnetized matter \cite{2010PhRvC..82f5802F, 2013NuPhA.898...43S}. 

Modified TOV equation can be derived after substitute Eqs. (\ref{metric}) and (\ref{total em tensor}) into Einstein field equation. The formula of modified TOV equation is 
\begin{equation}
    \frac{\mathrm{d}m}{\mathrm{d}r}=4\pi r^2\left(\rho + \frac{B^2}{8\pi} \right),\label{dmdr}
\end{equation}
\begin{equation}
    \frac{\mathrm{d}p_r}{\mathrm{d}r}=\begin{cases}
\frac{-\left(\rho+p_{r}\right) \frac{4 \pi r^{3}\left(p_{r}-\frac{B^{2}}{8 \pi}\right)+m}{r(r-2 m)}+\frac{2}{r} \Delta.}{1-\frac{\mathrm{d}}{\mathrm{d} \rho}\left(\frac{B^{2}}{8 \pi}\right)\left(\frac{\mathrm{d} \rho}{\mathrm{d} p_{r}}\right)}, & \text { for RO }, \\
\frac{-\left(\rho+p_{r}+\frac{B^{2}}{4 \pi}\right) \frac{4 \pi r^{3}\left(p_{r}+\frac{B^{2}}{8 \pi}\right)+m}{r(r-2 m)}+\frac{2}{r} \Delta}{1+\frac{\mathrm{d}}{\mathrm{d} \rho}\left(\frac{B^{2}}{8 \pi}\right)\left(\frac{\mathrm{d} \rho}{\mathrm{d} p_{r}}\right)}, & \text { for TO }.
\end{cases}\label{dpdr}
\end{equation}
The derivative of $p_r$ is separated by dominated orientation of magnetic field. RO represents radial orientation dominated while TO represent transerve orientation dominated. $m$ denote the mass and $\Delta$ is effective anisotropy of stellar structure with 
\begin{equation}
    \Delta = \begin{cases}
        p_t-p_r+\frac{B^2}{4\pi}, &\text{for RO},\\
        p_t-p_r-\frac{B^2}{8\pi}, &\text{for TO}.
    \end{cases}
\end{equation}
To numerically solve the modified TOV equations, we must specify the functional form of $\Delta$. Following \cite{2021ApJ...922..149D}, we adopt the expression 
\begin{equation}
    \Delta=
\begin{cases}
\kappa\frac{(\rho+p_r)\left(\rho+3p_r-\frac{B^2}{4\pi}\right)}{\left(1-\frac{2m}{r}\right)}r^2, & \mathrm{for~RO}, \\
 \\
\kappa\frac{\left(\rho+p_r+\frac{B^2}{4\pi}\right)\left(\rho+3p_r+\frac{B^2}{2\pi}\right)}{\left(1-\frac{2m}{r}\right)}r^2, & \mathrm{for~TO},
\end{cases}\label{anisotropy}
\end{equation}
where $\kappa$ is a dimensionless constant controlling the relative strength of anisotropy. Three assumptions of this model are summerized as follows \cite{2021ApJ...922..149D}:
\begin{enumerate}
    \item The anisotropy vanishes at the stellar center to ensure stability.
    \item The anisotropy varies with raius and is relative to $p_r$.
    \item The expression of anisotropy includes contributions arising from the local fluid and the magnetic field (magnitude and orientation).
\end{enumerate}

The values of $\kappa$ is within the range of $\left[-\frac23, \frac23\right]$ to ensure a credible solution of modified TOV. However, $p_t$ may turns negative over a narrow radial interval when $\kappa<0$ \cite{2019PhRvD..99j4002B}, which is consider as unphysical. Consequently, we restrict $\kappa$ within $\left[0, \frac23\right]$ to avoid unphysical results. 

Although many studies have attempted to construct the interior magnetic field of neutron star, the explicit distribution remains unknown. The magnetic field inside neutron stars may vary with density, hence a density-dependent model is suitable for our work. Following the \cite{1997PhRvL..79.2176B, 1998JPhG...24.1647B}, we adopt the parameteric model of magnetic field, which reads
\begin{equation}
    B(\rho)=B_s+B_0\left[1-\exp\left(-\eta\left(\frac{\rho}{\rho_0}\right)^\gamma\right)\right],\label{magnetic field}
\end{equation}
where $B_s$ is strength of surface magnetic field, $\eta$ and $\gamma$ dimensionless constants that control the \textbf{variation} of the field with density, $\rho_0$ is saturation density, and $B_0$ is a parameter with the same dimensions as the magnetic-field strength. Throughout this work we restrict to field strengths $B\leq3\times10^{18}~{\rm G}$ , for which Landau quantization effects are negligible (significant only for $B\geq10^{19}~{\rm G}$ \cite{2013NuPhA.898...43S}). We therefore neglect Landau quantization in subsequent calculations. 


\begin{figure}[b]
\includegraphics[scale=0.55]{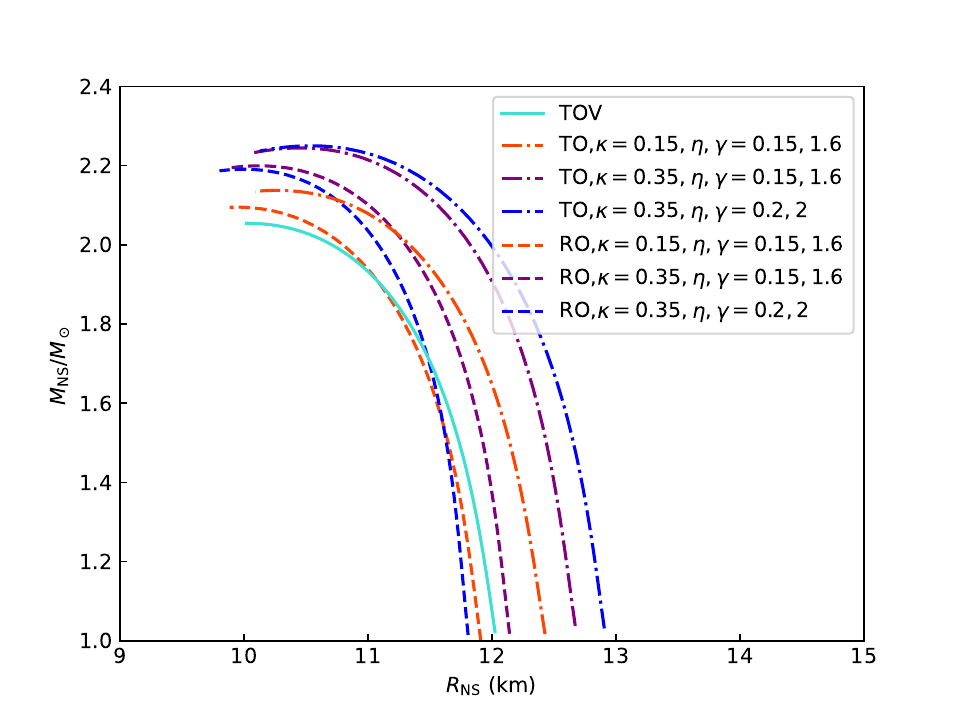}
\caption{\label{fig:1} $M_{\rm NS}-R_{\rm NS}$ curves for different parameter combinations. The solid line represents the solution of TOV eqaution. The dash line and dash dotted line represent the RO magnetic field and TO magnetic field, respectively.  }
\end{figure}

\subsection{Tidal Deformability}
Tidal deformability is the linear response of a neutron star to an external gravitational tidal field, $Q_{ij}=-\lambda \epsilon_{ij}$, where $Q_{ij}$ is quadrupole moment and $\epsilon_{ij}$ is static external quadrupolar tidal field. $\lambda$ is a constant which can be calculated by $\lambda = \frac23k_2R_{\rm NS}^{5}$ \cite{2008ApJ...677.1216H}, where $k_2$ is love number. In the analysis of GW, we are primarily interested in dimensionless tidal deformability $\Lambda=\lambda/M_{\rm NS}^5$. Unless otherwise specified, we use 'tidal deformability' to refer to the dimensionless tidal deformability $\Lambda$ throughout the remainder of this paper. 

$k_2$ need to be derived through solving the static, linearized perturbation equation arising from the external tidal field. The perturbated metric is
\begin{equation}
    g_{\mu\nu}=g_{\mu\nu}^{(0)} + h_{\mu\nu},
\end{equation}
where $g_{\mu\nu}^{(0)}$ is defined by Eq. (\ref{metric}). In the Regge–Wheeler gauge, the static, even-parity $l=2$ metric perturbation $h_{\mu\nu}$ can be written as \cite{1957PhRv..108.1063R, 1967ApJ...149..591T}
\begin{equation}
    \begin{aligned}
h_{\alpha \beta}= & \operatorname{diag}\left[e^{-\nu(r)} H_0(r), e^{\lambda(r)} H_{2}(r),\right. \\
& \left.r^{2} K(r), r^{2} \sin ^{2} \theta K(r)\right] Y_{2 m}(\theta, \phi),
\end{aligned}
\end{equation}
where $H_0$, $H_2$ and $K$ are functions of the radial coordinate $r$. The oridnary differential equation of $H(r)$ ($H_0=H_2\equiv H$) can be obtained by expanding the perturbated metric in Einstein field equation with anisotropy energy-momentum tensor, which can be expressed as \cite{2019PhRvD..99j4002B}
\begin{equation}
    \begin{array}{l}
H^{\prime \prime}+H^{\prime}\left[\frac{2}{r}+e^{\lambda}\left(\frac{2 m(r)}{r^{2}}+4 \pi r\left(p_{r}-\rho\right)\right)\right] \\
\quad+H\left[4 \pi e^{\lambda}\left(4 \rho+8 p_{r}+\frac{\rho+p_{r}}{A c_{s}^{2}}\left(1+c_{s}^{2}\right)\right)-\frac{6 e^{\lambda}}{r^{2}}-\nu^{\prime 2}\right]=0.
\end{array}\label{H}
\end{equation}
$c_s=(\partial p_r/\partial \rho)^\frac12$ denotes the sound speed alongside the radial direction. $e^\lambda=(1-2m/r)^{-1}$. $A=\mathrm{d}p_t/\mathrm{d}p_r$ and $\nu^\prime= 2 e^\lambda (m + 4\pi p_r r^3)/r^2$. For isotropical cases, $A=1$.

After solving the Eq. (\ref{H}), love number $k_2$ can be calculated by \cite{2008ApJ...677.1216H}
\begin{equation}
    \begin{aligned}k_2= & \frac85C^5(1-2C)^2[2-y+2C(y-1)]\\
& \times\{2C(6-3y+3C(5y-8))+4C^3\\
& \times[13-11y+C(3y-2)+2C^2(1+y)]\\
& + 3(1-2C)^2[2-y+2C(y-1)]\ln(1-2C)\}^{-1},\end{aligned}
\end{equation}
where $y=R_{\rm NS}H^\prime(R_{\rm NS})/H(R_{\rm NS})$ and $C=M/R$ is compactness of neutron star. 


\begin{figure}[b]
\includegraphics[scale=0.55]{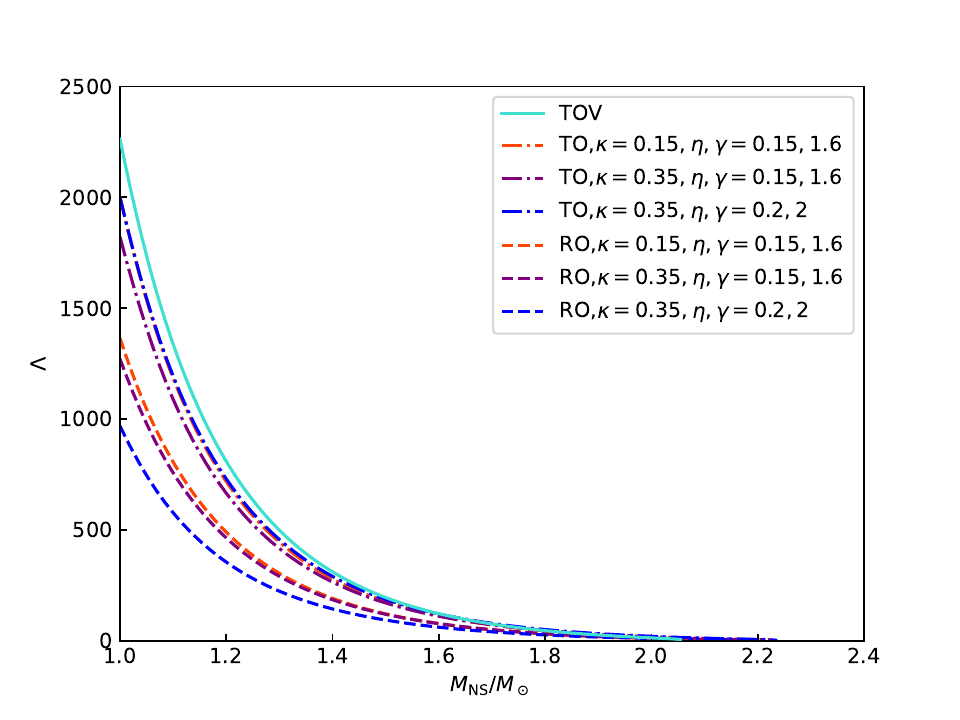}
\caption{\label{fig:2} $M_{\rm NS}-\Lambda$ curves for different parameter combinations. The solid line represents the solution of TOV eqaution. The dash line and dash dotted line represent the RO magnetic field and TO magnetic field, respectively.}
\end{figure}

\subsection{Gravitational wave}

To derive the solution of neutron star, we need EoS and together with the parameters $\kappa,~\eta,~\gamma$ and $B_0$. We analyze four mass combinations for the binary. One of the combination is a $1.27M_\odot$ neutron star merging with a $1.46M_\odot$ neutron star, taken from the GW170817 analysis \cite{2017PhRvL.119p1101A}. The rest of those are $1.2M_\odot-1.2M_\odot$, $1.2M_\odot-1.8M_\odot$ and $1.6M_\odot-1.6M_\odot$. We adopt three magnetic field configurations, RO, TO, and a non-magnetized interior named as No B, for each combination. Only $\kappa$ plays a role in non-magnetized cases, hence we fix $\kappa=0.35$. In RO and TO, we set $\kappa=0.35,~\eta=0.15,~\gamma=1.6$ and $B_0=5\times10^{17}~{\rm G}$ as standard cases. In the remaining three cases with the same magnetic field orientation, we vary one parameter set at a time, taking $\kappa=0.15$, $\eta=0.2$ together with $\gamma=2$ and $B_0=10^{18}~{\rm G}$, respectively. 

Figure 1 exhibits the $M_{\rm NS}-R_{\rm NS}$ curves with different combinations of parameters. Compare to TOV solution, all solution of modeified TOV increase the maximum mass of neutron star, especially magnetic field with TO. At a radius of $\sim11{\rm km}$, the mass of a neutron star in the TO configuration is larger by a factor of $\sim1.05$ than in the RO configuration for the standard parameter set. Figure 2 shows the $M_{\rm NS}-\Lambda$ curves with different combinations of parameters. The difference between the two dominant orientations decreases with increasing mass. To quantify the spread in tidal deformability predictions at a given mass, we compute, for each fixed $M_{\rm NS}$, the variance of $\Lambda$ over all considered configurations. Denoting by $\Lambda_i(M_{\rm NS})$ the tidal deformability of the $i$-th model and by $\bar{\Lambda}(M_{\rm NS})$ their arithmetic mean,
$$
\bar{\Lambda}(M_{\rm NS}) = \frac{1}{N}\sum_{i=1}^N \Lambda_i(M_{\rm NS}),
$$
the RMS fractional deviation (from the TOV solution) is defined as
$$
\frac{\sigma_\Lambda(M_{\rm NS})}{\Lambda_{\rm TOV}(M_{\rm NS})}
= \sqrt{\frac{1}{N}\sum_{i=1}^N
\left(\frac{\Lambda_i(M_{\rm NS})-\Lambda_{\rm TOV}(M_{\rm NS})}{\Lambda_{\rm TOV}(M_{\rm NS})}\right)^2 }.
$$
 For our set of models, this quantity is 0.192, 0.082, and 0.035 at $M_{\rm NS}=1.2M_\odot$, $1.4M_\odot$, and $1.6M_\odot$, respectively. This trend suggests that the tidal deformabilities of extremely compact neutron stars tend to converge, and thus have only a limited ability to distinguish between different internal structures. 

In this paper, we use \texttt{PyCBC} \cite{2019PASP..131b4503B, 2024zndo..10473621N} with the waveform model \texttt{IMRPhenomPv2\_NRTidalv2} \cite{2017PhRvD..96l1501D, 2019PhRvD.100d4003D} to generate GW signals. The neutron star is treated as non-rotating in our simulations. Only four parameters, $M_1,~M_2,~\Lambda_1,~\Lambda_2$, affect the simulation results. Tidal deformability affects both the phase and amplitude of the GW signal in the frequency domain. Since we are only interested in the SNR threshold, the source luminosity is not relevant in our simulations, and we fix the luminosity distance to 40 Mpc.

\section{Prospects of distinguishability for GW Detection}
To evaluate whether GWs can reveal anisotropy and magnetic field, we determine the signal-to-noise ratio (SNR) threshold that enables us to distinguish the internal structure of neutron star (for simplicity, we refer to this quantity as SNR threshold). If the SNR threshold is within a reasonable range, we conclude that GWs can probe internal anisotropy and magnetic field. The defination of mismatch can be expressed as
\begin{equation}
    \mathcal{M}=1-\max_{t_c,\phi_c} \frac{(h_1(t_c,~\phi_c)| h_2)}{\sqrt{(h_1|h_1)(h_2|h_2)}}.
\end{equation}
$(h_1|h_2)$ is noise-weighted overlap, defined as
\begin{equation}
    (h_1|h_2)=4 \Re\int_{f_{\min }}^{f_{\max }} \frac{\tilde{h}_1(f) \tilde{h}^*_2(f)}{S_n(f)} \mathrm{d} f,
\end{equation}
where $\tilde{h}_1(f)$ denotes the GW strain in the frequency domain. $S_n(f)$ is the power spectral density (PSD) of the GW detector. In this paper, we use the O4 PSD \cite{LVK_T2200043}. $f_\mathrm{min}$ and $f_\mathrm{max}$ are $20~{\rm Hz}$ and $2048~{\rm Hz}$, respectively. 

The SNR threshold can be estimated as \cite{2017PhRvD..95j4004C}
\begin{equation}
    \mathrm{SNR_{th}}=\sqrt{\frac{D}{2\mathcal{M}}},
\end{equation}
where $D$ is the number of intrinsic model parameters.

We analyze four mass combinations for the binary. First combination is a $1.27M_\odot$ neutron star merging with a $1.46M_\odot$ neutron star, taken from the GW170817 analysis. The rest of those are $1.2M_\odot-1.2M_\odot$, $1.2M_\odot-1.8M_\odot$ and $1.6M_\odot-1.6M_\odot$. The chirp mass for each combinations is $1.1848M_\odot$, $1.0447M_\odot$, $1.2742M_\odot$ and $1.3929M_\odot$. The results are conclude in Table 1. 
\begin{table*}
\caption{\label{tab:table1}SNR threshold required to distinguish different interior magnetic-field orientations and anisotropy levels in the modified TOV framework. Rows labeled TO and RO denote transverse-orientation–dominated and radial-orientation–dominated internal fields, respectively. No B is the anisotropic, non-magnetized reference. $B_0^*=B_0/10^{18}~{\rm G}$. Columns list four representative binary neutron star mass pairs. Each table entry gives the required SNR threshold, followed in parentheses by the corresponding fractional deviation in the effective binary tidal deformability, $|(\tilde{\Lambda}_{\rm mod}-\tilde{\Lambda}_{\rm TOV})/\tilde{\Lambda}_{\rm TOV}|$. Across all cases, RO configurations yield lower SNR, represents easier to distinguish than TO at comparable $\kappa$ and field prescriptions.}
\begin{ruledtabular}
\begin{tabular}{ccccccc}
Orientation &  $\kappa$ & $\eta,~\gamma,~B_0^*$ & \multicolumn{4}{c}{SNR threshold($|(\tilde{\Lambda}_{\rm mod}-\tilde{\Lambda}_{\rm TOV})/\tilde{\Lambda}_{\rm TOV}|$)} \\
\cline{4-7}
 &  &  & $1.2M_\odot$--$1.2M_\odot$
 & $1.27M_\odot$--$1.46M_\odot$
 & $1.2M_\odot$--$1.8M_\odot$
 & $1.6M_\odot$--$1.6M_\odot$ \\ \hline
 TO & 0.35 & 0.15, 1.6, 0.5 & 48(0.18) & 78.04(0.17) & 83.02(0.18) & 213.73(0.09) \\
 & 0.15 & 0.15, 1.6, 0.5 & 71.64(0.12) & 117.6(0.11) & 125.08(0.12) & 357.63(0.07) \\
 & 0.35 & 0.20, 2.0, 0.5 & 81.26(0.10) & 150.92(0.09)  & 155(0.10) & 443.42(0.01) \\
 & 0.35 & 0.15, 1.6, 1.0 & 87.54(0.09) & 106.35(0.10) & 127.79(0.09) & 125.23(0.21)\\
 \hline
RO & 0.35 & 0.15, 1.6, 0.5 & 25.69(0.43) & 36.87(0.42) & 37.94(0.43) & 74.17(0.37)\\
 & 0.15 & 0.15, 1.6, 0.5 & 27.04(0.39) & 38.33(0.39) & 39.88(0.39) & 74.2(0.37) \\
 & 0.35 & 0.20, 2.0, 0.5 & 21.62(0.56) & 30.58(0.54) & 31.6(0.56) & 56.54(0.50)\\
 & 0.35 & 0.15, 1.6, 1.0 & 18.35(0.74) & 25.57(0.73) & 26.53(0.74) & 42.1(0.70)\\
\hline
No B & 0.35 & - & 36.6(0.16) & 55.73(0.15) & 58.62(0.15) & 155.54(0.17) \\
\end{tabular}
\end{ruledtabular}
\end{table*}
The SNR threshold increases with increasing chirp mass when the anisotropy and magnetic field parameters are held fixed across mass combinations. This trend arises because discrepancies in tidal deformability are more pronounced at lower masses. Consequently, analyses of mergers of lower mass neutron star binaries are most likely to reveal internal anisotropy and the magnetic field structure. Our results show that the dominant magnetic field orientation has a pronounced influence on the quantity of SNR threshold. If most neutron stars host RO magnetic fields, these effects could be revealed from GW analyses. This is due to the RO field acts as an effective central (radially inward) force, making the star more compact than in pure TOV configurations. By contrast, a TO field acts as an outward directed effective force, leading to a comparatively smaller change in tidal deformability. 

Since TO field is unlikely to be distinguishable at current sensitivities of GW detector, we focus exclusively on RO cases. In addition to the fact that lower mass neutron stars increase discrepancies in tidal deformability, which reduces the SNR threshold, the magnetic field distribution and magnitude also affect this threshold. A stronger magnetic field with a slower decay toward lower density lowers the SNR threshold, enabling GW observations to more effectively reveal the neutron star’s interior structure.

The mass distribution of Galactic binary neutron stars peaks at $1.3M_\odot$ \cite{2016ARA&A..54..401O, 2019ApJ...876...18F}, a regime where anisotropy and magnetic field effects are most pronounced. If most neutron stars host RO internal magnetic fields, and if the cosmic BNS mass distribution matches the Galactic one (peaking near $1.3M_\odot$), then GW analyses do have ability to reaveal the nature of interior. However, the structure of neutron star internal magnetic fields is still unclear, and which configuration (RO or TO) predominates remains debated. The internal magnetic field distribution is difficult to constrain, as there are currently no robust observational methods. We adopt a density-dependent model to circumvent the current theoretical uncertainties, although this choice may itself introduce biases. Since a first-principles equation for anisotropy is difficult to derive theoretically, we instead employ a phenomenological prescription. Given this uncertainty, more detailed models of anisotropic stars and a full treatment of the magnetic field distribution are needed in future work. The deployment of third generation GW detectors promises a glance of neutron star internal physics, enabled by superior instrumental sensitivity and a substantially expanded catalog of binary neutron star merger events. These results encourage us to consider more detailed models of anisotropy and magnetic field distributions to improve the accuracy of GW analyses, thereby enabling more accurate inference of the physical parameters of neutron star interiors. 

\section{conclusion}
In this paper, we consider a modified TOV equation that incorporates the effects of pressure anisotropy and magnetic fields. Neutron stars with anisotropy and magnetic fields can sustain higher maximum masses than in the isotropic, non-magnetized case, with anisotropy playing a more decisive role than the magnetic field. These effects modify GW signals primarily by changing the tidal deformability at fixed mass. Their imprints may already be marginally accessible with current instruments.

According to our results, when all other parameters are held fixed, GW analyses more easily identify anisotropic neutron stars with RO magnetic fields than those with TO fields. For example, a $1.2M_\odot-1.2M_\odot$ binary neutron star system with $\kappa=0.35$, $\eta=0.15$ and $\gamma=1.6$ needs at least ${\rm SNR}\simeq48$ to reveal its interior structure if the field is TO, but only $\simeq 26$ if the field is RO. We consider four pairs of component masses of binary neutron star merger system: $1.2M_\odot-1.2M_\odot$, $1.27M_\odot-1.46M_\odot$, $1.2M_\odot-1.8M_\odot$ and $1.6M_\odot-1.6M_\odot$ with chirp mass $1.0447M_\odot$, $1.18487M_\odot$, $1.2742M_\odot$ and $1.3929M_\odot$, respectively. As shown in Table \ref{tab:table1}, the SNR threshold decreasing as chirp mass increasing, indicating that lower chirp mass makes the interior structure easier to identified from GW data. We also find that the SNR threshold depends differently on the orientation of magnetic field. As the magnetic field strength increases in the TO case, the SNR threshold rises from 78 to 105, making it harder to investigate the interior physics from GW analysis with current detectors. On the contrary, the SNR threshold decreases from 36 to 25 if magnetic field is RO. This tendency arises from different effect for TO and RO magnetic field, which reduces compactness of neutron star for TO whereas RO steepen it. Overall, current GW analyses are more likely to reveal neutron stars structures with RO magnetic field. This provides a new opportunity to investigate the structure of neutron stars \cite{2025arXiv251122987W}.

\begin{acknowledgments}
We thank Quan Cheng and  Guo-Peng Li for helpful discussions.
\end{acknowledgments}

\bibliography{apssamp.bbl}

\end{document}